\documentclass{article}
\usepackage{ijcai25}

\usepackage{times}
\usepackage{soul}
\usepackage{url}
\usepackage[hidelinks]{hyperref}
\usepackage[utf8]{inputenc}
\usepackage[small]{caption}
\usepackage{graphicx}
\usepackage{amsmath}
\usepackage{amsthm}
\usepackage{booktabs}
\usepackage{algorithm}
\usepackage{algorithmic}
\usepackage[switch]{lineno}

\usepackage{amsfonts}
\usepackage{multirow}

\title{HyperDet: Source Detection in Hypergraphs via Interactive Relationship Construction and Feature-rich Attention Fusion}

\author{
Le Cheng$^{1,2}$\and
Peican Zhu$^1$\thanks{Corresponding authors.}\and
Yangming Guo$^3$\and
Keke Tang$^4$\footnotemark[1]\and
Chao Gao$^1$\And
Zhen Wang$^{3}$\footnotemark[1]\\
\affiliations
$^1$School of Artificial Intelligence, Optics and Electronics, Northwestern Polytechnical University\\
$^2$School of Computer Science, Northwestern Polytechnical University\\
$^3$School of Cybersecurity, Northwestern Polytechnical University\\
$^4$Cyberspace Institute of Advanced Technology, Guangzhou University\\
\emails
\{ericcan, w-zhen\}@nwpu.edu.cn,
tangbohutbh@gmail.com
}

\begin{document}

\maketitle

\begin{abstract}
Hypergraphs offer superior modeling capabilities for social networks, particularly in capturing group phenomena that extend beyond pairwise interactions in rumor propagation. Existing approaches in rumor source detection predominantly focus on dyadic interactions, which inadequately address the complexity of more intricate relational structures. In this study, we present a novel approach for Source Detection in Hypergraphs (HyperDet) via Interactive Relationship Construction and Feature-rich Attention Fusion. Specifically, our methodology employs an Interactive Relationship Construction module to accurately model both the static topology and dynamic interactions among users, followed by the Feature-rich Attention Fusion module, which autonomously learns node features and discriminates between nodes using a self-attention mechanism, thereby effectively learning node representations under the framework of accurately modeled higher-order relationships. Extensive experimental validation confirms the efficacy of our HyperDet approach, showcasing its superiority relative to current state-of-the-art methods.
\end{abstract}

\begin{figure}[htbp]
  \centering
  \includegraphics[width=\linewidth]{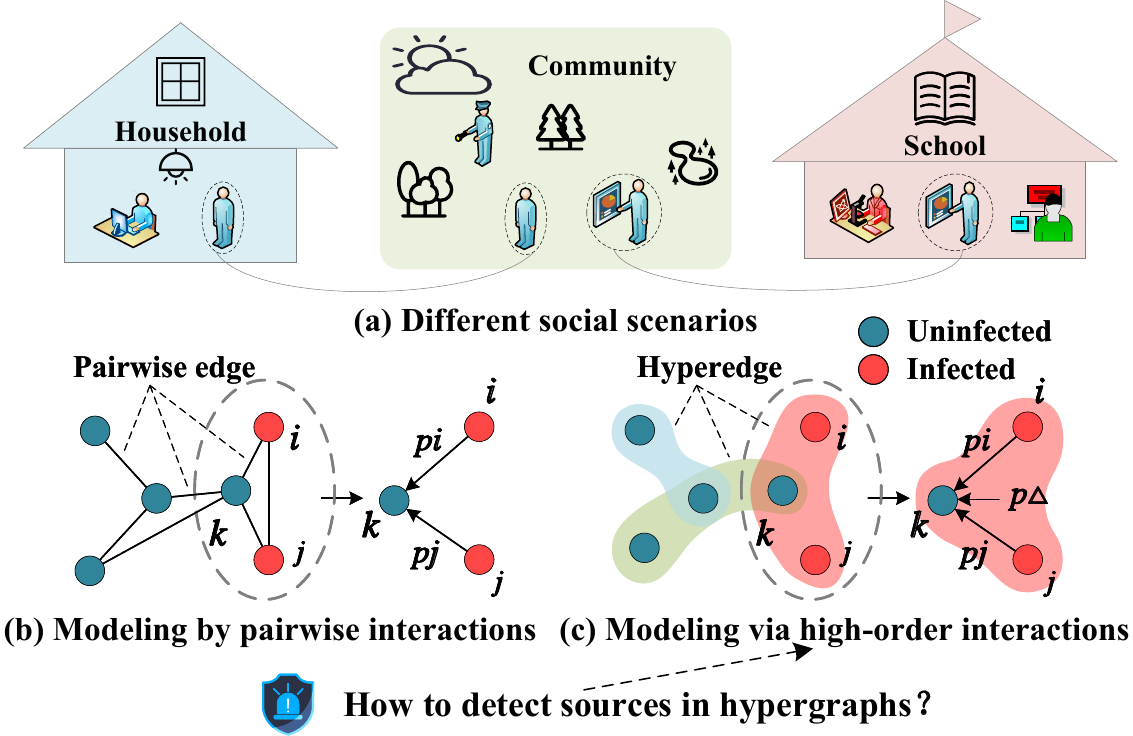}
  \caption{Social scenario modeling. (a) Various social scenarios. (b) Pairwise interactions in exiting methods. (c) Higher-order interactions in our proposed approach. This accurate yet complex hypergraph modeling introduces new challenges to source detection.}
  \label{fig1}
\end{figure}

\section{Introduction}
Source detection in graphs offers a viable mathematical approach to identifying the sources of propagation information such as rumors \cite{shah2011rumors,zhu2022locating}. Over recent years, various methodologies have been proposed to address this issue \cite{wang2025elevating,dong2024unveiling}. Initially, techniques based on source centrality theory \cite{shah2011rumors} and maximum likelihood estimation \cite{yang2020locating}, including methods such as LPSI \cite{wang2017multiple}, OJC \cite{zhu2017catch}, and MLE \cite{pinto2012locating,cheng2025efficient}. More recently, with the rapid advancement of graph neural networks \cite{kipf2016semi,jin2023local}, an increasing number of researchers have employed strategies to learn node representations by embedding node features and modeling propagation information \cite{ling2022source,wang2022invertible}. These approaches have set benchmarks in identifying sources \cite{cheng2024heuristic}.

However, existing methods predominantly model graphs through dyadic interactions, assuming that user interactions are pairwise. In practice, interactions often extend beyond pairwise engagements, occurring among triads or larger groups, thus constituting higher-order interactions \cite{battiston2020networks,gao2022hgnn+}. To our knowledge, no existing work has addressed source detection considering interactions that extend beyond the pairwise level. When networks exhibit higher-order interactions, the influence/pressure among group peers accelerates the information propagation process, in stark contrast to binary interactions, as illustrated in Fig. \ref{fig1}.

Hypergraphs offer a superior framework for modeling the aforementioned higher-order interactions. However, source detection in hypergraphs faces two significant challenges. First, propagation on hypergraphs exhibit both static low/high-order topological structures and dynamic user interactions, posing a primary challenge in accurate modeling. Second, node features are often manually engineered and of low dimensionality, which can hinder model performance. Furthermore, the varying significance of different nodes in information propagation, if treated uniformly, affects the model's learning and convergence. To intuitively address these issues, we should: 1) Establishing relationships between nodes through both static topology and dynamic interactions to enhance subsequent message propagation. 2) Enabling the model to autonomously learn and enrich node features, while distinguishing between nodes based on their information transmission capabilities.

In this paper, we propose a novel Source Detection approach in Hypergraphs (HyperDet) via Interactive Relationship Construction and Feature-rich Attention Fusion. Initially, to accurately depict user relationships and node features, an Interactive Relationship Construction module (IRC) is developed to leverage the static topology of the hypergraph as well as the dynamic interactions between infected and uninfected subgraphs to form the nodes' interactive relationships; combining node states, propagation information, and positional encoding to constitute raw node features. Subsequently, to automatically learn and enrich node features, and to focus on nodes with stronger information propagation capabilities, a Feature-rich Attention Fusion module (FAF) is introduced. This module includes a designed hypergraph autoencoder to extract latent node features, and a multi-head attention mechanism to automatically assign different attention coefficients to various nodes. Lastly, to counteract model prediction biases due to significant sample size differences between sources and non-sources, we introduce a class balancing mechanism. The effectiveness of our proposed method is validated across eight public datasets, and extensive experiments demonstrate that HyperDet outperforms state-of-the-art methods in hypergraph-based source detection.

Overall, our contributions can be summarized as follows:
\begin{itemize}
    \item We are the first to formalize the problem of source detection in hypergraphs and to propose a methodology to address this challenge.
    \item We introduce a novel approach for source detection in hypergraphs through Interactive Relationship Construction and Feature-rich Attention Fusion.
    \item We demonstrate the effectiveness of our proposed method over baseline approaches through extensive experimental validation.
\end{itemize}

\section{Related Work}
\paragraph{General Source Detection.}
Various methods have been proposed for source detection \cite{bao2024graph}. Based on the source centrality theory \cite{prakash2012spotting,shah2011rumors}, LPSI selects local outliers by iterating node labels \cite{wang2017multiple}, EPA computes the infection time for each node and selects the node with the longest duration of infection \cite{ali2019epa}, and OJC identifies nodes that cover the infected area with the smallest radius \cite{zhu2017catch}. However, these methods do not account for user diversity and have limited application scenarios. 

\paragraph{Graph Learning-based Source Detection.}
Utilizing graph neural networks (GNNs), GCNSI and SIGN employ user states as inputs to classify nodes and identify sources \cite{dong2019multiple,li2021propagation}, while GCSSI considers the last infected node, referred to as the wavefront \cite{dong2022wavefront}. From the perspective of network structure, ResGCN enhances the influence of initial features through a residual structure \cite{shah2020finding}. Nevertheless, these methods are primarily challenged by class imbalance and lack integration with the information propagation process. 

Addressing graph diffusion explicitly, IVGD initially learns the information propagation process and subsequently reverses this knowledge for source detection \cite{wang2022invertible}. SL-VAE deciphers various propagation patterns on graphs \cite{ling2022source}, and GIN-SD accommodates nodes with partial information \cite{cheng2024gin}. Despite these advances, the above methods predominantly model social network propagation through pairwise interactions. This simplification overlooks higher-order interactions that are prevalent in group dynamics, a crucial aspect that remains unaddressed by these models \cite{battiston2020networks}.

\paragraph{Hypergraph Neural Networks.}
To model higher-order social phenomena, hypergraphs have been proposed with hyperedges containing multiple nodes beyond mere pairwise connections \cite{zhou2006learning,battiston2020networks}. Building on the higher-order structure of hypergraphs, HGNN introduces a two-stage message passing approach to learn node representations, consisting of hyperedge information aggregation and node information aggregation \cite{feng2019hypergraph,gao2022hgnn+}. Inspired by the Graph Attention Network \cite{velivckovic2017graph}, Hyper-Atten enhances the model's learning capability by computing attention coefficients for nodes and their associated hyperedges \cite{bai2021hypergraph}. 

Based on GNNs, UniGNN directly applies GNNs to hypergraphs using hyperedges as intermediaries \cite{huang2021unignn}, while HyperGCN first transforms hypergraphs into weighted graphs before applying GCNs for node representation learning \cite{yadati2019hypergcn}. Integrating existing approaches, AllSet proposes an effective model that unifies current hypergraph learning methods \cite{chien2021you}. As an extension of simple pairwise graphs, hypergraphs demonstrate a stronger and more flexible data modeling capability, widely applied and exhibiting high efficiency across multiple domains \cite{jiao2024enhancing}.


\begin{figure*}[t]
  \centering
  \includegraphics[width=\linewidth]{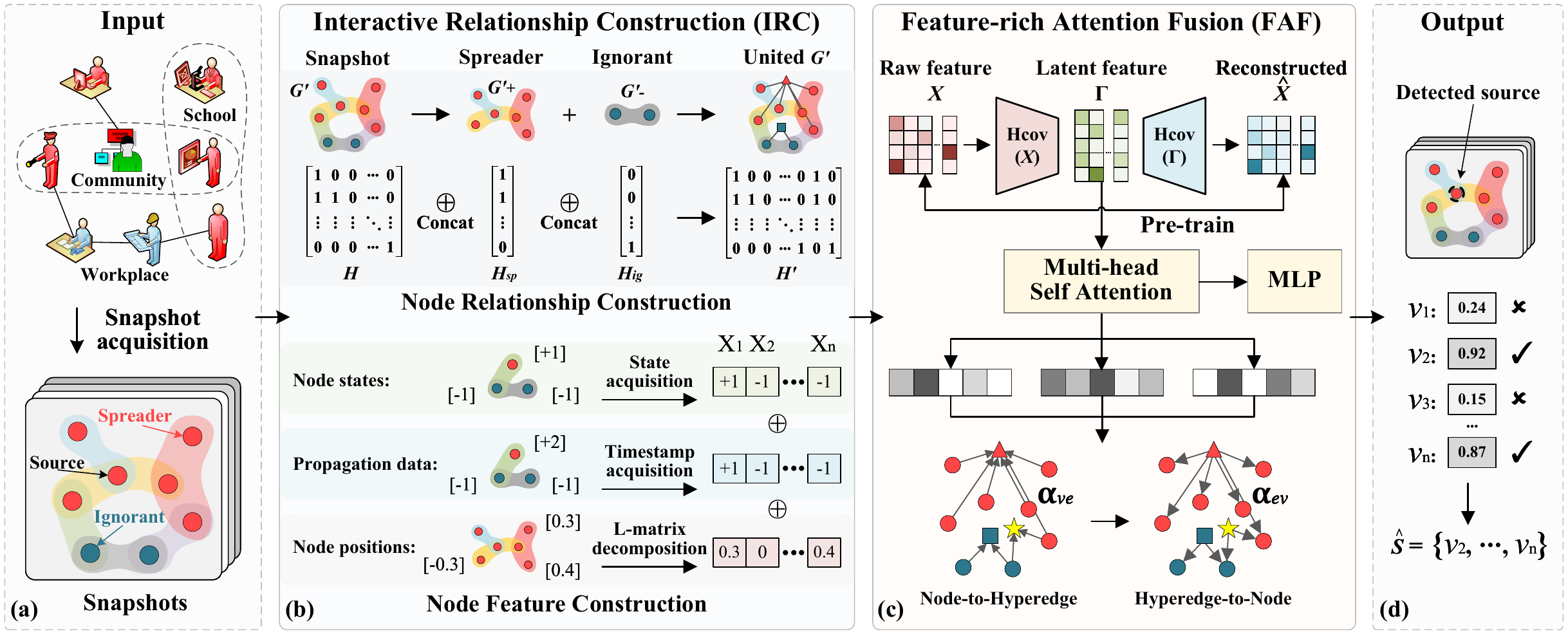}
  \vspace{-4mm}
  \caption{Framework of HyperDet. (a) The snapshot acquisition process. (b) The Interactive Relationship Construction (IRC) module, which constructs node relationships through static topology and dynamic interactions, and initially builds raw node features. (c) The Feature-rich Attention Fusion (FAF) module, which first learns node representations via autoencoding and then differentiates nodes using a self-attention mechanism to enhance model learning capabilities. (d) Model output, where nodes are classified based on predicted probabilities to identify the source set $\hat{s}$.}
  \label{framework}
  \vspace{-3mm}
\end{figure*}

\section{Problem Formulation}
\paragraph{Preliminary on Hypergraphs.}
The low/high-order interactions in social networks can be modeled using a hypergraph $G$ = $(V,E,\mathbf{\Omega})$, where $V$ = $\{v_1, v_2, ...,v_n\}$ represents the set of nodes; $E$ = $\{e_1,e_2,...,e_m\}$ denotes the set of hyperedges, with each hyperedge $e_i$ = $\{v_{i1},v_{i2},...,v_{ij}\}$, where $i \in \{1,2,...,m\}$ and $j \in \{1,2,...,n\}$; and $\mathbf{\Omega} \in \mathbb{R}^{m \times m}$ is a diagonal matrix, each element of which represents the weight of the corresponding hyperedge. Based on the relationship between node $v$ and hyperedge $e$, the hypergraph can be represented by an incidence matrix $\textit{\textbf{H}} \in \mathbb{R}^{n \times m}$, i.e.,
\begin{align}
\textbf{\textit{H}}_{ve} & = \left\{\begin{array}{ll}
1, & v \in e \\
0, & \text {otherwise}.
\end{array}\right.
\end{align}

Consequently, the degree of node $v$ in the hypergraph can be defined as $\textit{\textbf{D}}_{vv}$ = $\sum_{e=1}^m \mathbf{\Omega}_{ee} \textit{\textbf{H}}_{ve}$, and the degree of hyperedge $e$ can be defined as $\textit{\textbf{D}}_{ee}$ = $\sum_{v=1}^n \textit{\textbf{H}}_{ve}$, where the diagonal matrices $\textit{\textbf{D}}_V \in \mathbb{R}^{n \times n}$ and $\textit{\textbf{D}}_E \in \mathbb{R}^{m \times m}$ respectively represent the degree matrices for nodes and hyperedges.

\paragraph{Propagation Process on Hypergraphs.}
The information propagation on hypergraphs unfolds over the evolutionary time $t$, where at $t$ = 0, the set of source nodes $s$ in the network transitions from the ignorant (susceptible) state to the spreader (infected) state. As time progresses, each spreader $v_i$ disseminates information to its neighbors with a probability $p_i$. Unlike the propagation processes on bipartite graphs, the propagation on hypergraphs not only involves pairwise interactions but also includes peer influence/pressure within groups represented by hyperedges. Consequently, an ignorant node transitions to the spreader state with an additional probability $p_\triangle$, as depicted in Fig. \ref{fig1}. Several classical models, such as the IC, SI, SIR, and SIS models \cite{battiston2020networks,de2020social}, have been proposed to characterize the propagation process on hypergraphs. 

In summary, the propagation process on hypergraphs can be represented by the time series $\{G'(t),t \geq 0\}$, where $G'(t)$ denotes the network snapshot at time $t$, with nodes being categorized into two types based on the states, i.e., $G_+$ (spreader) and $G_-$ (ignorant).

\paragraph{Source Detection in Hypergraphs.}
As the propagation proceeds, when $\delta\%$ of the nodes in the network reach the spreader state, a network snapshot $G'$ is captured, which includes the network's topological structure $T$, the states of the nodes $N$, and the propagation information $P$. Consequently, source detection in a hypergraph can be formalized as: 
\begin{equation}
\hat{s}=f(G'(T,N,P)),
\end{equation}
where $f(\cdot)$ represents the corresponding source detection method, and $\hat{s}$ denotes the detected source set.

\paragraph{Discussion.}
For source detection in hypergraphs, the network snapshot $G'$ not only encapsulates static low/high-order topological structures but also contains dynamic user interaction information, underscoring the necessity to accurately model the relationships between nodes. Furthermore, the node features are manually engineered with low dimensionality and the attributes of nodes are diverse; thus, enabling the model to automatically learn, enrich these node features, and differentiate between nodes becomes crucial for enhancing model performance.

\section{Method}
In this section, we introduce the proposed HyperDet, which primarily consists of two modules: the Interactive Relationship Construction module (IRC) and the Feature-rich Attention Fusion module (FAF), as illustrated in Fig. \ref{framework}.


\subsection{Interactive Relationship Construction (IRC)}
The IRC module comprises two submodules, i.e., Node Relationship Construction and Node Feature Construction.
\paragraph{Node Relationship Construction.}
The hypergraph snapshot $G'$ incorporates not only the static topological structure but also the dynamic interaction information among nodes, we model these two aspects using hyperedges. For the static topological structure, it is represented by the incidence matrix $\textit{\textbf{H}}$, where each hyperedge $e$ encompasses nodes that can interact with each other. Regarding the dynamic interaction information, nodes in $G'$ that are in ignorant and spreader states are encompassed within specific hyperedges, denoted as $e_{ig}$ = $\{v_i | v_i \in G_-\}$ and $e_{sp}$ = $\{v_j | v_j \in G_+\}$, respectively, which can further be transformed into the incidence matrices $\textit{\textbf{H}}_{ig} \in \{0,1\}^{n \times 1}$ and $\textit{\textbf{H}}_{sp} \in \{0,1\}^{n \times 1}$. Therefore, the overall incidence matrix can be concatenated as:
\begin{equation}
\textit{\textbf{H}}' \in \{0,1\}^{n \times (m+2)} = \textit{\textbf{H}}~\|~ \textit{\textbf{H}}_{ig}~\|~\textit{\textbf{H}}_{sp}.
\label{con_H}
\end{equation}

\paragraph{Node Feature Construction.}
Various types of information, including node states, propagation data, and node positions, are embedded as node features.

The hypergraph snapshot $G'(t)$ captures the state of node $v_i$ at time $t$ within the network: spreader ($v_i \in G_+$), indicating that $v_i$ is influenced by and propagating the rumor, and ignorant ($v_i \in G_-$), meaning $v_i$ has not received the rumor or deems the information unreliable. Consequently, the state feature $\textit{\textbf{X}}_i^1$ can be represented as:
\begin{equation}
\textbf{\textit{X}}_i^1 = \left\{\begin{array}{ll}
+1, & v_i \in G_+ \\
-1, & \text {otherwise}.
\end{array}\right.
\end{equation}

Social platforms such as Facebook and Twitter record timestamps when users forward information, which are critical for understanding the propagation dynamics. Therefore, for node $v_i$, we incorporate the information timestamp as its propagation features:
\begin{equation}
\textbf{\textit{X}}_i^2 = \left\{\begin{array}{ll}
\ \ t_i, & v_i \in G_+ \\
-1, & \text {otherwise}.
\end{array}\right.
\end{equation}

In spatial domain convolution, unlike the message passing aggregated from local neighbors, the positional relationships between nodes play a crucial role in facilitating global message propagation. Additionally, encoding the different positions of infected nodes assists the model in learning the locational characteristics of the source. Given the generative nature of Laplacian positional encoding, we decompose the Laplacian matrix of the infected subgraph for encoding the positional features of nodes.




We first extract the infected subgraph $G_+'$ from $G'$, and its symmetric normalized Laplacian matrix is computed as:
\begin{equation}
\textit{\textbf{L}}^{sym}_+ = \textit{\textbf{I}} - \textit{\textbf{D}}^{-\frac{1}{2}}_{V+} \textit{\textbf{H}}_+ \textit{\textbf{D}}^{-1}_{E+} \textit{\textbf{H}}_+^T \textit{\textbf{D}}^{-\frac{1}{2}}_{V+},
\end{equation}
where $\textit{\textbf{H}}_{+}$ is the incidence matrix of $G_+'$, $\textit{\textbf{D}}_{V+}$ and $\textit{\textbf{D}}_{E+}$ respectively represent the corresponding degree matrices of the nodes and edges. Subsequently, the matrix $\textit{\textbf{L}}_+^{sym}$ is subjected to factorization:
\begin{equation}
\bigtriangleup_{\textit{\textbf{L}}_+^{sym}} = \mathbf{\Psi} \mathbf{\Lambda} \mathbf{\Psi}^T, 
\end{equation}
here, $\mathbf{\Lambda}$ is a diagonal matrix whose elements are eigenvalues, while matrix $\mathbf{\Psi}$ consists of the corresponding eigenvectors. Given that the first component of the eigenvectors weakly differentiates between node structure and organizational information, we select $k$ smallest non-trivial eigenvectors starting from the second component as $\mathbf{\Psi}_i$ for node $v_i$'s positional encoding $(k \ll n)$. Therefore, the positional feature of node $v_i$ is represented as:
\begin{equation}
\textbf{\textit{X}}_i^3 = \left\{\begin{array}{ll}
\mathbf{\Psi}_i, & v_i \in G_+' \\
-1, & \text {otherwise}.
\end{array}\right.
\label{X_3}
\end{equation}

In conclusion, the raw feature vector of node $v_i$ can be synthesized by concatenating several component features to achieve a comprehensive representation:
\begin{equation}
\textbf{\textit{X}}_i = \left[\|_{x=1}^3 \textbf{\textit{X}}_i^x\right].
\end{equation}

\subsection{Feature-rich Attention Fusion (FAF)}
The FAF module first automates the learning and enrichment of raw node features, and then differentiates between nodes using a self-attention mechanism.


\paragraph{Feature Augmentation via Autoencoding.}
In contemporary graph-based learning paradigms, the manually engineered raw node features may not adequately support model learning and stability due to their simplistic representation and limited discriminative power. To address this, we design an autoencoder with encoder-decoder architecture. Specifically, the feature vector $\textit{\textbf{X}}$ undergoes transformation by the encoder En$(\cdot)$, resulting in an embedding $\mathbf{\Gamma}$, which is subsequently reconstructed back into $\hat{\textit{\textbf{X}}}$ by the decoder De$(\cdot)$. This process is encapsulated as:
\begin{equation}
    \hat{\textit{\textbf{X}}} = \operatorname{De}\left(\operatorname{En}\left( \textit{\textbf{X}}, \textit{\textbf{H}}' \right) , \textit{\textbf{H}}' \right).
\end{equation}


Each layer of this architecture involves successive transformations through hypergraph convolution and attention mechanism $\operatorname{HConv_{att}}(\cdot)$. The process can be formalized as:
\begin{equation}
    \textit{\textbf{X}}^{(l+1)} = \operatorname{LReLU}\left( \operatorname{HConv_{att}}\left( \textit{\textbf{X}}^{(l)}, \mathbf{\Theta}^{(l)}, \textit{\textbf{H}}', \mathbf{\Omega} \right) \right),
\end{equation}
where $\textbf{\textit{X}}^{(l)} \in \mathbb{R}^{l_w \times n}$ represents the features of nodes at layer $l$ and $\textit{\textbf{X}}^{(0)}$ = $\textit{\textbf{X}}$, $\mathbf{\Theta}^{(l)} \in \mathbb{R}^{l_w \times m}$ denotes the hyperedge attributes at the $l^{th}$ layer, $l_w$ is the dimensionality of both node features and hyperedge attributes. $\textit{\textbf{H}}'$ and $\mathbf{\Omega}$ are the incidence matrix and hyperedge weights matrix, respectively. $\operatorname{LReLU}(\cdot)$ serves as the activation function. The attention coefficients $\alpha$ are computed based on the relative importance of nodes in relation to their associated hyperedges:
\begin{equation}
\alpha_{i j} = \frac{\exp \left(\vec{a}^{T}\operatorname{LReLU}\left(\textbf{\textit{W}}\left[ \textbf{\textit{X}}_{i} \| \mathbf{\theta}_{j}\right]\right)\right)}{\sum_{e_k \in \mathcal{N}(v_i)} \exp \left(\vec{a}^{T}\operatorname{LReLU}\left(\textbf{\textit{W}}\left[ \textbf{\textit{X}}_{i} \| \mathbf{\theta}_{k}\right]\right)\right)},
\label{aij}
\end{equation}
where $\Vec{a} \in \mathbb{R}^{2l_w'}$ represents a learnable parameter of the attention mechanism, $\textbf{\textit{W}} \in \mathbb{R}^{l_w' \times l_w}$ is the weight matrix, and $\mathcal{N}(v_i)$ denotes the set of hyperedges that include node $v_i$ within the hypergraph. $\textit{\textbf{X}}_i$ and $\mathbf{\theta}_j$ represent the features of node $v_i$ and the attributes of hyperedge $e_j$, respectively.

Building on the aforementioned architectures, each layer of hypergraph attention convolution is delineated into two distinct operations, i.e., 

\textbf{(1) Node-to-Hyperedge Convolution}
In this process, each node transmits its features to the connected hyperedges:
\begin{equation}
\mathbf{\theta}_{e}^{\prime} = \sum_{v \in \mathcal{N}(e)} \alpha_{v e} \textbf{\textit{W}} \textbf{\textit{X}}_{v},
\label{edge_singlehead}
\end{equation}
where $\mathcal{N}(e)$ represents the set of nodes being connected to the hyperedge $e$.

\textbf{(2) Hyperedge-to-Node Convolution}
Conversely, in the hyperedge-to-node convolution, the features of hyperedges are aggregated back to the nodes:
\begin{equation}
\textit{\textbf{X}}_{v}' = \sum_{e \in \mathcal{N}(v)} \alpha_{e v} \textbf{\textit{W}}' \mathbf{\theta}_{e}^{\prime}.
\label{node_singlehead}
\end{equation}

To facilitate effective learning, the autoencoder is initially pre-trained by minimizing the discrepancy between the original node features and the reconstructed features, fostering an accurate representation of the latent space conducive to downstream tasks. The loss function utilized during this pre-training phase is defined as:
\begin{equation}
    \mathcal{L}_{ae} = \| \textit{\textbf{X}} - \hat{\textit{\textbf{X}}} \|^2.
\end{equation}

\paragraph{Differential Attention Fusion.}
The latent node representations $\mathbf{\Gamma}$ learned by the autoencoder often require nuanced processing due to the diversity of nodes. To address this, we employ hypergraph attention convolution to allocate distinct attention coefficients to different nodes. The transformation for $\mathbf{\Gamma}$ is formalized by:
\begin{equation}
    \mathbf{\Gamma}^{\prime} = \operatorname{LReLU}\left( \operatorname{HConv_{att}}\left( \mathbf{\Gamma}, \mathbf{\Theta}, \textit{\textbf{H}}', \mathbf{\Omega} \right) \right),
    \label{att_first}
\end{equation}
where $\textit{\textbf{H}}'$ is the incidence matrix constructed in Eq. (\ref{con_H}) within the IRC module.

Furthermore, to enhance the discriminative power of our model, a multi-head attention mechanism is employed. This approach involves the fusion of $K$ independent attention computations to refine the node representations, which are then concatenated to serve as the input for the subsequent layer. The process is formalized as:
\begin{equation}
   \mathbf{\Gamma}^{\prime\prime} = {\|}_{k = 1}^K \operatorname{LReLU}\left(\operatorname{HConv_{att}}\left( \mathbf{\Gamma}^{\prime k} \right)\right).
   \label{multihead}
\end{equation}

In the final layer of attention module, to align dimensions and synthesize the information across different heads, the representations from $K$ attention heads are averaged:
\vspace{-2mm}
\begin{equation}
\mathbf{\Gamma}^{\prime\prime\prime} = \frac{1}{K} \sum_{k = 1}^{K} \mathbf{\Gamma}^{\prime\prime k}.
\end{equation}

This operation ensures the final node representation encapsulates a comprehensive view, mitigating the bias that might arise from a single convolution channel. Ultimately, these node representations are projected into a 2-dimensional space and transformed via a softmax function to calculate the probabilities of each node belonging to the source/non-source set:
\begin{equation}
P(v_x) = \frac{e^{z_{i}}}{\sum_{j} e^{z_{j}}},\  \Vec{z} = \mathbf{\Gamma}_{x}^{\prime\prime\prime}.
\end{equation}

This sophisticated architecture leverages both the detailed local interactions and the global structure of the hypergraphs, thereby ensuring that our model effectively discerns the nuanced roles of nodes in information propagation.

\subsection{Loss Function and Training}
To address the significant disparity in sample sizes between the source node set and the non-source set, thereby mitigating prediction bias, we design a balancing coefficient:
\begin{equation}
\rho = \frac{|s|}{n-|s|},
\end{equation}
where $n$ and $|s|$ represent the number of nodes and sources, respectively. This approach ensures that the contribution of each node is weighted equally. Integrating this class balance mechanism, we configure the loss function for the attention fusion module as:
\begin{equation}
\mathcal{L}_{af} = \sum_{v_{i} \in s} \mathcal{L}_{i}+\rho \sum_{v_{j} \in(V-s)} \mathcal{L}_{j},
\label{loss}
\end{equation}
here, $\mathcal{L}$ denotes the cross-entropy loss, for sample $x$ and its label $y$, $\mathcal{L}(x,y)$ = $-y\log(\hat{y})$, $\hat{y}$ is the predicted probability.

During the training process, we fine-tune the autoencoder, and the overall loss function is formulated as:
\begin{equation}
    \mathcal{L}_{HyperDet} = \mathcal{L}_{ae} + \mathcal{L}_{af} + \lambda\|w\|_{2},
\end{equation}
where $\lambda$ is the regularization parameter, and $\| \cdot \|^2$ represents the L2-norm, adding a penalty for model complexity to prevent overfitting.

For the proposed HyperDet system, the IRC module establishes node relationships and features. Subsequently, the FAF module undertakes the automatic learning and enrichment of node features, followed by the application of multi-head attention to differentiate nodes based on their significance and roles within the network. This structured approach ensures effective identification of source nodes within diverse and intricate network structures.

\section{Experiments}
\subsection{Experimental Settings}
\paragraph{Implementation.}
Due to the independent attributes of users in social networks, the short-term nature and the group effects of rumor propagation, we employ a heterogeneous independent cascade model on hypergraphs to simulate the spread of rumors. Initially, 5$\%$ of the nodes are selected as sources, with each node’s propagation probability $p$ following a uniform distribution $U(0,0.5)$. The group propagation probability $p_\triangle$ is proportional to the ratio of infected nodes within a hyperedge, represented as $p_\triangle$ = $0.3 (|e \cap G_+|/|e|)$. A network snapshot is captured when 30$\%$ of the nodes become spreaders, and the final ratio of the training set to the test set is 8:2. For the learning process, the autoencoder and the HyperDet system are set with learning rates of 0.01 and 0.005, respectively, with the latent node feature dimensions at 64. For small networks ($G_1$-$G_4$), there are three attention heads and 500 hidden neurons; medium networks ($G_5$-$G_7$) use two heads and 500 neurons; and the large network ($G_8$) has one head and 400 neurons due to space constraints. The hypergraphs undergo clique expansion before being applied in the baseline methods. All experiments are conducted on a workstation equipped with four NVIDIA RTX 3090 GPUs.

\paragraph{Datasets.}
We evaluate the methods using eight diverse and widely utilized datasets across various scales and domains, including Zoo \cite{asuncion2007uci}, House \cite{chodrow2021generative}, NTU2012 \cite{chen2003visual}, Mushroom \cite{asuncion2007uci}, ModelNet40 \cite{wu20153d}, 20News \cite{asuncion2007uci}, PubMed \cite{yadati2019hypergcn}, and Walmart \cite{amburg2020clustering}.



\begin{table*}[!h]
\centering
\resizebox{0.98\textwidth}{!}{%
\begin{tabular}{l|ccc|ccc|ccc|ccc}
\hline
\multirow{2}{*}{\textbf{Methods}} & \multicolumn{3}{c|}{\textbf{Zoo}}  & \multicolumn{3}{c|}{\textbf{House}}      & \multicolumn{3}{c|}{\textbf{NTU2012}}  & \multicolumn{3}{c}{\textbf{Mushroom}}  \\ \cline{2-13}
 & \textbf{ACC} & \textbf{F-Score} & \textbf{AUC} &  \textbf{ACC} & \textbf{F-Score} & \textbf{AUC} & \textbf{ACC} & \textbf{F-Score} & \textbf{AUC} & \textbf{ACC} & \textbf{F-Score} & \textbf{AUC} \\ \hline
LPSI & 0.798 & 0.348 & 0.785 & 0.806 & 0.341 & 0.804 & 0.795 & 0.316 & 0.823 & 0.810 & 0.332 & 0.780 \\
EPA & 0.793 & 0.356 & 0.790 & 0.812 & 0.336 & 0.812 & 0.806 & 0.302 & 0.815 & 0.816 & 0.329 & 0.764 \\
GCNSI & 0.819 & 0.378 & 0.805 & 0.826 & 0.322 & 0.809 & 0.814 & 0.297 & 0.829 & 0.809 & 0.358 & 0.798 \\
SIGN & 0.824 & 0.457 & 0.816 & 0.829 & 0.417 & 0.810 & 0.819 & 0.408 & 0.826 & 0.811 & 0.436 & 0.820 \\
GCSSI & 0.817 & 0.418 & 0.801 & 0.815 & 0.397 & 0.799 & 0.813 & 0.384 & 0.825 & 0.810 & 0.395 & 0.810 \\
ResGCN & 0.829 & 0.497 & 0.810 & 0.824 & 0.459 & 0.814 & 0.805 & 0.439 & 0.829 & 0.821 & 0.447 & 0.826 \\
IVGD & 0.835 & 0.578 & 0.809 & 0.815 & 0.513 & 0.829 & 0.852 & 0.526 & 0.841 & 0.842 & 0.510 & 0.858 \\
SL-VAE & 0.834 & 0.598 & 0.815 & 0.841 & 0.528 & 0.815 & 0.840 & 0.517 & 0.832 & 0.852 & 0.526 & 0.884 \\
GIN-SD & 0.856 & 0.682 & 0.849 & 0.853 & 0.617 & 0.847 & 0.866 & 0.629 & 0.850 & 0.861 & 0.631 & 0.878 \\
HyperDet (ours) & \textbf{0.901} & \textbf{0.792} & \textbf{0.892} & \textbf{0.884} & \textbf{0.726} & \textbf{0.875} & \textbf{0.877} & \textbf{0.715} & \textbf{0.869} & \textbf{0.912} & \textbf{0.801} & \textbf{0.923} \\ \hline

\multirow{2}{*}{\textbf{Methods}} & \multicolumn{3}{c|}{\textbf{ModelNet40}}  & \multicolumn{3}{c|}{\textbf{20News}}      & \multicolumn{3}{c|}{\textbf{PubMed}}  & \multicolumn{3}{c}{\textbf{Walmart}}  \\ \cline{2-13}
 & \textbf{ACC} & \textbf{F-Score} & \textbf{AUC} &  \textbf{ACC} & \textbf{F-Score} & \textbf{AUC} & \textbf{ACC} & \textbf{F-Score} & \textbf{AUC} & \textbf{ACC} & \textbf{F-Score} & \textbf{AUC} \\ \hline
LPSI & 0.764 & 0.235 & 0.786 & 0.732 & 0.255 & 0.758 & 0.749 & 0.181 & 0.762 & 0.705 & 0.125 & 0.712 \\
EPA & 0.785 & 0.249 & 0.791 & 0.756 & 0.267 & 0.764 & 0.762 & 0.192 & 0.778 & 0.710 & 0.114 & 0.725 \\
GCNSI & 0.810 & 0.254 & 0.801 & 0.795 & 0.218 & 0.785 & 0.805 & 0.188 & 0.789 & 0.806 & 0.180 & 0.799 \\
SIGN & 0.820 & 0.429 & 0.818 & 0.809 & 0.359 & 0.796 & 0.804 & 0.229 & 0.784 & 0.819 & 0.236 & 0.805 \\
GCSSI & 0.808 & 0.378 & 0.786 & 0.813 & 0.357 & 0.782 & 0.803 & 0.189 & 0.786 & 0.796 & 0.195 & 0.795 \\
ResGCN & 0.814 & 0.482 & 0.813 & 0.820 & 0.426 & 0.811 & 0.806 & 0.235 & 0.817 & 0.819 & 0.224 & 0.818 \\
IVGD & 0.826 & 0.610 & 0.823 & 0.821 & 0.534 & 0.827 & 0.824 & 0.531 & 0.819 & 0.815 & 0.523 & 0.824 \\
SL-VAE & 0.836 & 0.579 & 0.824 & 0.835 & 0.561 & 0.819 & 0.818 & 0.530 & 0.819 & 0.829 & 0.514 & 0.836 \\
GIN-SD & 0.826 & 0.607 & 0.816 & 0.839 & 0.580 & 0.821 & 0.834 & 0.574 & 0.826 & 0.829 & 0.527 & 0.836 \\
HyperDet (ours) & \textbf{0.887} & \textbf{0.786} & \textbf{0.885} & \textbf{0.869} & \textbf{0.748} & \textbf{0.882} & \textbf{0.870} & \textbf{0.723} & \textbf{0.871} & \textbf{0.868} & \textbf{0.627} & \textbf{0.859} \\ \hline
\end{tabular}%
}
\caption{The performance of source detection across all datasets for each method, with the best results highlighted in bold.}
\label{overallperformance1}
\vspace{-2mm}
\end{table*}

\paragraph{Evaluation Metrics.}

Accuracy (ACC), F1-Score, and Area Under the Curve (AUC) are used to assess method performance. ACC measures the correct classification rate of samples. The F1-Score comprises Precision, which quantifies the ratio of true sources in the predicted set $\hat{s}$, calculated as $|\hat{s} \cap s|\ /\ |\hat{s}|$, and Recall, which assesses the proportion of true sources correctly detected, computed as $|\hat{s} \cap s|\ /\ |s|$. AUC evaluates the model’s capacity to differentiate between source and non-source classes at all threshold levels. These metrics provide a comprehensive framework for assessing the effectiveness and reliability of source detection methods.

\paragraph{Baselines.}
Several representative methods that published recent years are considered as baselines, including those based on source centrality such as LPSI \cite{wang2017multiple} and EPA \cite{ali2019epa}; methods that account for user states like GCNSI \cite{dong2019multiple}, SIGN \cite{li2021propagation}, GCSSI \cite{dong2022wavefront} and ResGCN \cite{shah2020finding}; and approaches integrating propagation information such as IVGD \cite{wang2022invertible}, SL-VAE \cite{ling2022source}, and GIN-SD \cite{cheng2024gin}.

\vspace{-1mm}
\subsection{Performance Analyses}
\vspace{-1mm}
\paragraph{Comparison with State-of-the-art Methods.}
The comparative results are compiled in Table \ref{overallperformance1}. From the outcomes, we can discern several insights: 1) All methods exhibit relatively high ACC, while F-Scores are generally lower. This discrepancy primarily arises because ACC accounts for all samples, including non-sources, thus a greater divergence between ACC and F-Score indicates a more severe impact of class imbalance issues on the model, as observed with the LPSI, EPA, and GCNSI methods. 2) Methods that integrate propagation information, such as IVGD, SL-VAE, and GIN-SD, demonstrate superiority over those based solely on source centrality theory or user states. This advantage is largely due to the inherent randomness of the propagation process, which cannot be effectively countered by merely considering user states. Overall, in the context of all evaluated methods, the presence of higher-order interactions means that the propagation process is not confined to pairwise interactions, making it challenging for baseline methods based on pairwise interactions to capture propagation dynamics effectively. Consequently, these methods underperform compared to their optimal outcomes. However, the proposed HyperDet achieves the best results across all datasets. Specifically, HyperDet improves performance by 23$\%$-30$\%$ over methods that based on source centrality theory and user states, and by 8$\%$-15$\%$ over those consider propagation information. This improvement is attributable primarily to HyperDet's strategies: 1) Designing higher-order interaction relationships among nodes through both static topology and dynamic interactions. 2) Enriching node features and discriminatively learning node representations through autoencoding and multi-head attention mechanisms. 3) Eliminating predictive bias in the model through a class balancing mechanism.

\paragraph{Visualization.}
To provide an intuitive representation of source detection results, we visualize the correctly detected sources of representative methods on the House network, as shown in Fig. 3. This visualization facilitates an immediate understanding of the comparative performance and effectiveness of each method in a real-world network context.

\begin{figure}[htbp]
  \centering
  \includegraphics[width=\linewidth]{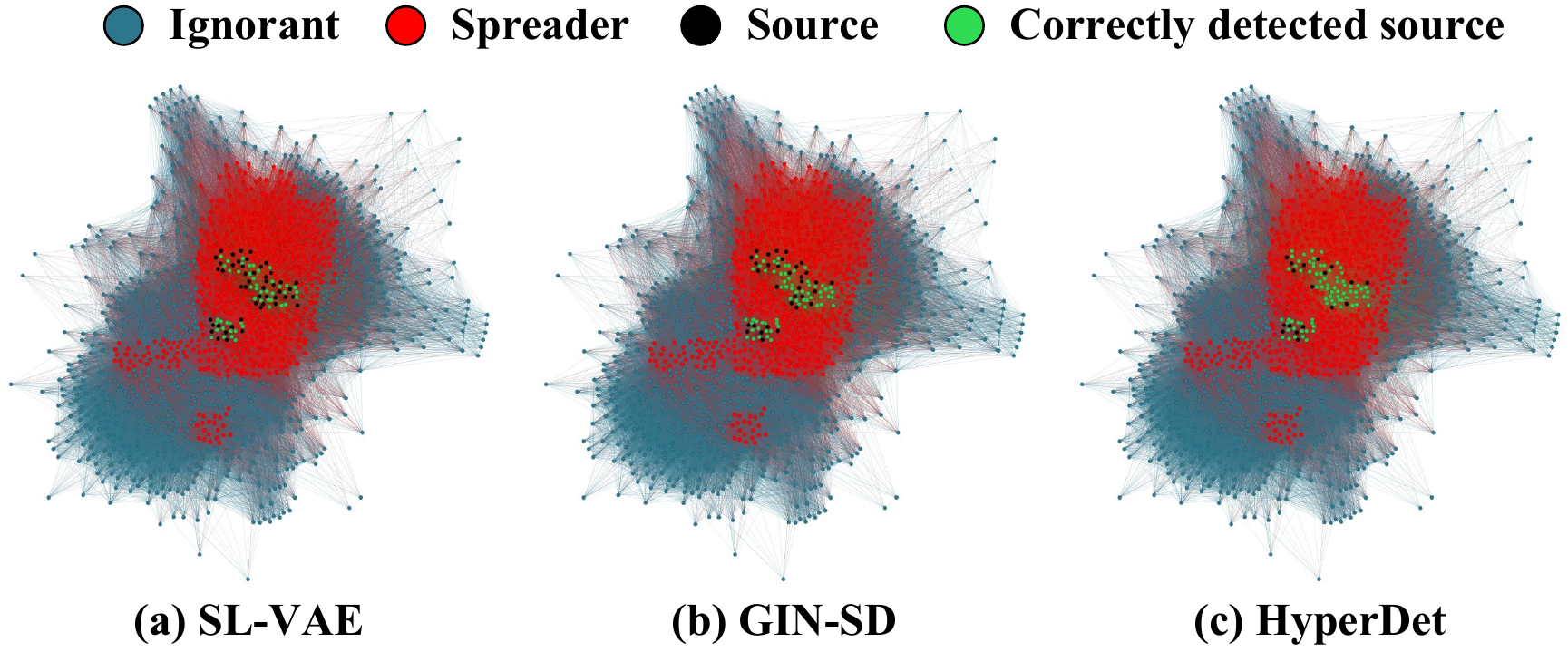}
  \vspace{-5mm}
  \caption{Visualization of source detection results on House.}
  \label{early_detection}
  \vspace{-5mm}
\end{figure}

\paragraph{Performance on Early Detection.}
Early detection of rumor sources is crucial to prevent further escalation of harm. We analyze the performance of various methods as the scale of rumor propagation increases from 10$\%$ to 30$\%$, in an increment of 5$\%$. The results are summarized in Fig. \ref{early_detection}. As the scale of propagation expands, the accuracy of all methods gradually declines. This trend is primarily due to the increasing number of infected nodes that need to be differentiated, underscoring the importance of early intervention in rumor source detection. Moreover, HyperDet outperforms baseline methods across all datasets, demonstrating its efficiency.

\begin{figure}[tbp]
  \centering
  \includegraphics[width=\linewidth]{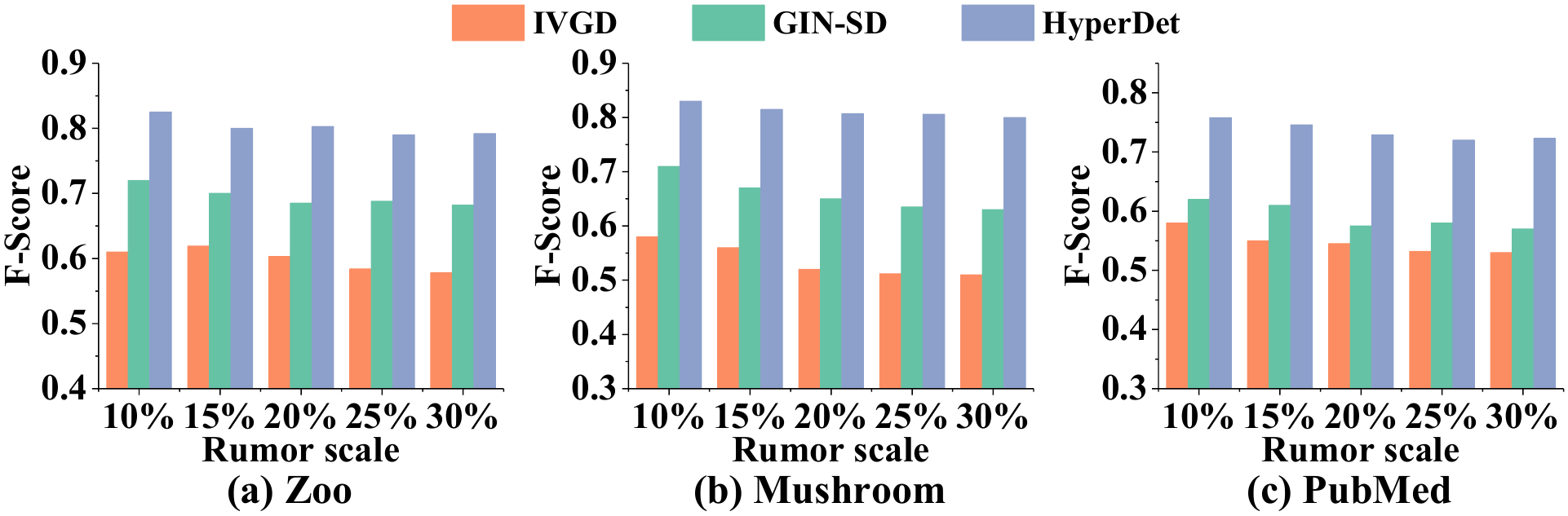}
  \vspace{-5mm}
  \caption{The performance in early rumor sources detection.}
  \label{early_detection}
\vspace{-3mm}  
\end{figure}

\paragraph{Impact of Data Incompleteness.}

\begin{figure}[htbp]
  \centering
  \includegraphics[width=\linewidth]{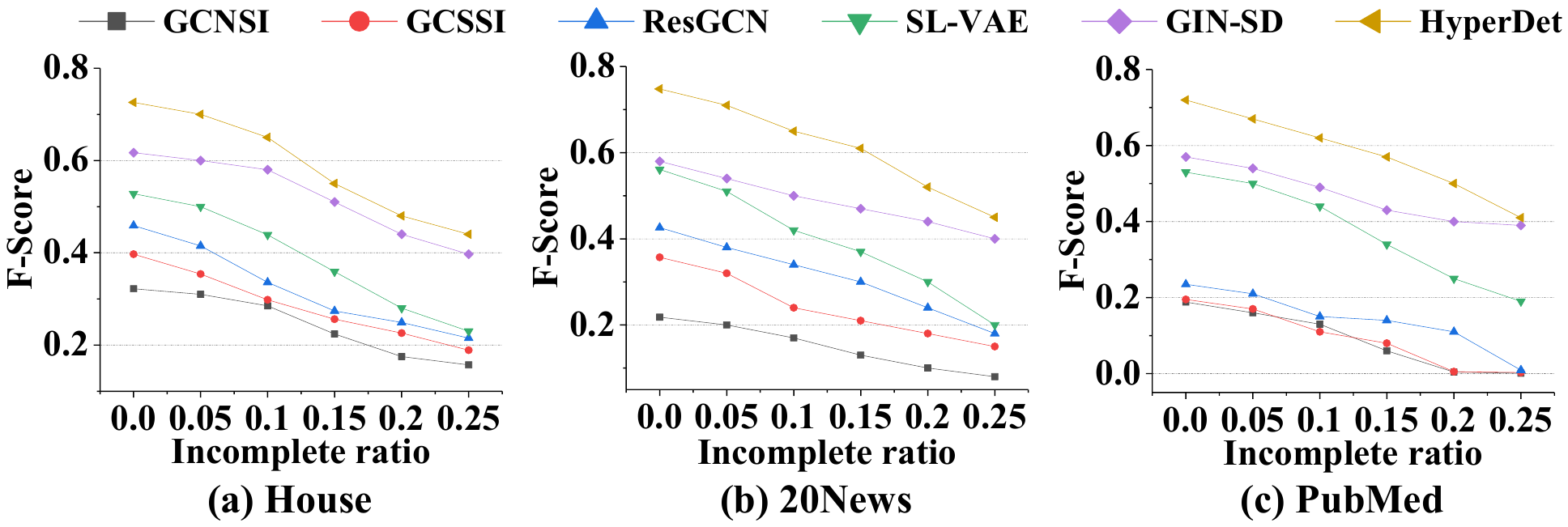}
  \vspace{-5mm}
  \caption{Impact of data incompleteness on source detection.}
  \label{Incomplete_ratio}
\vspace{-4mm}
\end{figure}

The data incomplete rates are increased from 0 to 0.25 in steps of 0.05, indicating that a corresponding proportion of nodes lack features. The results are shown in Fig. \ref{Incomplete_ratio}. As the proportion of incomplete data increases, there is a noticeable decline in the performance of all methods. This decrease is primarily due to the incomplete features of nodes, which hinder the model's ability to effectively learn and converge. However, the performance of HyperDet decreases the least, which is mainly because in addition to node states and propagation features, we also consider structural features, specifically positional encodings. This integration diminishes the impact of feature loss on the model's performance, demonstrating the robustness and adaptability of the proposed HyperDet.

\paragraph{Computational Efficiency.}
The runtime of HyperDet compared to methods that integrate propagation information is detailed in Table \ref{time}. Due to the requirement for additional modules to learn propagation dynamics, methods such as IVGD and SL-VAE exhibit longer runtimes. In contrast, HyperDet incorporates propagation information into feature embeddings, thereby reducing model complexity. This demonstrates its enhanced efficiency.
\begin{table}[htbp]
\centering
\resizebox{0.71\linewidth}{!}{%
\begin{tabular}{lccc}
\hline
\textbf{Datasets} & \textbf{IVGD}  & \textbf{SL-VAE} & \textbf{HyperDet} \\  \hline
NTU2012 & 179.71 & 198.53 & \textbf{126.11}  \\
Mushroom & 264.30 & 295.05 & \textbf{212.86} \\ \hline
\end{tabular}%
}
\caption{The runtime (s) comparison across different methods on NTU2012 and Mushroom datasets.}
\label{time}
\vspace{-5mm}
\end{table}

\subsection{Ablation Study and Other Analyses}


\begin{table}[htbp]
\centering
\resizebox{0.96\linewidth}{!}{%
\begin{tabular}{c|l|cc|cc}
\hline
\multirow{2}{*}{\textbf{Module}} & \multirow{2}{*}{\textbf{Methods}} & \multicolumn{2}{c|}{\textbf{ModelNet40}}  & \multicolumn{2}{c}{\textbf{PubMed}} \\ \cline{3-6}
 & & \textbf{ACC} & \textbf{F-Score} & \textbf{ACC} & \textbf{F-Score} \\ \hline
\multirow{2}{*}{IRC}
& w/o H & 0.835 & 0.629 & 0.840 & 0.611  \\
& w/o D & 0.854 & 0.753 & 0.847 & 0.705  \\ \hline
\multirow{4}{*}{FAF}
& w/o E & 0.874 & 0.771 & 0.852 & 0.705 \\
& w/o A & 0.837 & 0.739 & 0.846 & 0.697 \\
& w/ AL & 0.825 & 0.731 & 0.839 & 0.689 \\
& w/ AS & 0.774 & 0.358 & 0.781 & 0.260 \\ \hline
- & HyperDet & \textbf{0.887} & \textbf{0.786} & \textbf{0.870} & \textbf{0.723} \\ \hline
\end{tabular}%
}
\caption{Performance of different HyperDet variants.}
\label{ablation}
\vspace{-4mm}
\end{table}

\paragraph{Effect of Interactive Relationship Construction.}
We initially remove the high-order structure from the hypergraph; results from w/o H indicate a significant performance decline compared to configurations that retain this structure, as shown in Table \ref{ablation}. Additionally, by preserving the high-order topology while eliminating the dynamic interaction construction, as in w/o D, we observe an improvement over w/o H, yet performance remains inferior to the full HyperDet model. These findings underscore the importance of both static topological structures and dynamic interactions in capturing the high-order relationships between nodes.

\paragraph{Effect of Feature-rich Attention Fusion.}
Removing the autoencoder, as indicted in Table \ref{ablation}, w/o E demonstrates that manually engineered features with sparse dimensions degrade model performance. Regarding attention, we design three variants: treating nodes and hyperedges uniformly, w/o A achieves the best results. w/ AL is next, primarily due to the presence of small-degree yet crucial bridge nodes. w/ AS performs the worst, mainly because it overly focuses on nodes with lesser information transmission capabilities. These findings emphasize the importance of automatic learning and enrichment of features, as well as adaptive attention in enhancing model efficacy.



\paragraph{Effects of Different Information Diffusion Models.}
The performance of HyperDet under various propagation models is depicted in Table \ref{models}. Experimental results affirm that HyperDet consistently exhibits stable performance not only in the influence-based IC model but also in the infection-based SI, SIS, and SIR models. Notably, in the SIS and SIR models, the sources may randomly recover and thus become obscured, leading to a decline in performance. These findings substantiate the versatility and broad applicability of HyperDet across diverse modeling scenarios.

\begin{table}[htbp]
\centering
\resizebox{\linewidth}{!}{%
\begin{tabular}{c|ccc|ccc}
\hline
\multirow{2}{*}{\textbf{Models}} & \multicolumn{3}{c|}{\textbf{House}}  & \multicolumn{3}{c}{\textbf{NTU2012}} \\ \cline{2-7}
 & \textbf{ACC} & \textbf{F-Score} & \textbf{AUC} & \textbf{ACC} & \textbf{F-Score} & \textbf{AUC}\\ \hline
SI & 0.878 & 0.728 & 0.881 & 0.879 & 0.711 & 0.873 \\
SIS & 0.863 & 0.692 & 0.851 & 0.859 & 0.685 & 0.862 \\
SIR & 0.854 & 0.687 & 0.850 & 0.862 & 0.680 & 0.865 \\
IC & 0.884 & 0.726 & 0.875 & 0.877 & 0.715 & 0.869 \\ \hline
\end{tabular}%
}
\caption{The performance of HyperDet under different information propagation models.}
\label{models}
\vspace{-3mm}
\end{table}
\vspace{-2mm}

\section{Conclusion}
This paper explores the phenomenon of group interactions within social networks and introduces a novel source detection approach, HyperDet, applied to hypergraphs to address this challenge. The key idea lies in integrating static topology with dynamic interactions to construct node relationships, followed by enhancing model learning capabilities through feature enrichment and adaptive attention mechanisms. Extensive comparative and ablation experiments on hypergraphs demonstrate HyperDet's efficacy and robustness against state-of-the-art methods. We hope this work can inspire source detection researchers and advances the development of effective modeling in this field.

\section*{Acknowledgements}
This work was supported in part by the National Natural Science Foundation of China (Grant Nos. U22B2036, 62025602, 62471403, 62073263, 62472117), the Fundamental Research Funds for the Central Universities (Grant Nos. G2024WD0151, D5000240309), the Technological InnovationTeam of Shaanxi Province (Grant No. 2025RS-CXTD-009), the International Cooperation Project of Shaanxi Province (Grant No. 2025GH-YBXM-017), the Guangdong Basic and Applied Basic Research Foundation (Grant No. 2025A1515010157), the Science and Technology Projects in Guangzhou (Grant No. 2025A03J0137),  the National Foreign Experts Program (Grant No. G2023183019), and the Tencent Foundation and XPLORER PRIZE.


\bibliographystyle{named}
\bibliography{ijcai25}

\end{document}